\documentclass[prl,superscriptaddress,twocolumn,showpacs,%
amsmath,amssymb]{revtex4-1}

\usepackage{slashed}
\usepackage{graphicx}
\usepackage{color}

\newcommand{\bea}{\begin{eqnarray}}
\newcommand{\eea}{\end{eqnarray}}
\newcommand{\beq}{\begin{equation}}
\newcommand{\eeq}{\end{equation}}

\begin{document}
\title{Geometrically Induced Magnetic Catalysis and Critical Dimensions}

\author{Antonino Flachi}
\affiliation{Multidisciplinary Center for Astrophysics and
 Department of Physics, Instituto Superior T\'{e}cnico,
 University of Lisbon, Avenida Rovisco Pais 1,  1049-001 Lisboa, Portugal}

\author{Kenji Fukushima}
\affiliation{Department of Physics, The University of Tokyo,
  7-3-1 Hongo, Bunkyo-ku, Tokyo 113-0033, Japan}

\author{Vincenzo Vitagliano}
\affiliation{Multidisciplinary Center for Astrophysics \& Department of Physics, Instituto Superior T\'{e}cnico,
University of Lisbon, Avenida Rovisco Pais 1,  1049-001 Lisboa, Portugal}

\begin{abstract}
 We discuss the combined effect of magnetic fields and geometry in
 interacting fermionic systems.  
{At leading order in the heat-kernel expansion, the infrared singularity (that in flat space leads to the magnetic catalysis) is regulated by the chiral gap effect, and the catalysis is deactivated by the effect of the scalar curvature. We discover that an infrared singularity is found in higher-order terms that mix the magnetic field with curvature, and these lead to a novel form of geometrically induced magnetic catalysis.}
The dynamical mass squared is then modified not only due to the chiral gap effect by an amount proportional to the curvature, 
 but also by a magnetic shift $\propto (4-D)eB$, where $D$ represents the number of space-time
 dimensions.  We argue that $D=4$ is a critical dimension across which
 the behavior of the magnetic shift changes qualitatively.
\end{abstract}
\pacs{04.62.+v, 12.38.Aw, 90.80.-k}
\maketitle

%%%%%%%%%%   Introduction   %%%%%%%%%%
%\paragraph{Introduction:}
The phenomenon of spontaneous symmetry breaking, originally inspired
by the BCS theory of superconductivity, was initially established to
account for the masses of nucleons and light pions \cite{njl}.  In
modern terminology, we say that chiral symmetry, i.e., a symmetry
between left- and right-handed sectors of massless fermions, is
spontaneously broken by a nonvanishing chiral condensate
$\langle\bar{\psi}\psi\rangle\neq 0$.  When this happens, a dynamical
mass for the fermions, denoted here as $M$, is generated by such a
condensation phenomenon.  Since the early days, theorists have been
trying to develop ways to compute the effective potential as a function
of the chiral condensate in order to be able to unfold the nature of
symmetry breaking and of its associated phase transitions in a wide
variety of contexts.

The chiral condensate plays an essential role in the characterization
of the phase structure of quark matter in extreme conditions, for
instance, at high temperature or density, as well as in the presence of
external (magnetic, gravitational, etc.) fields (see
Refs.~\cite{Fukushima:2011jc,magnreview,Inagaki:1997kz} for reviews).
However, theoretical applications are not limited to quark matter, and
concepts and methods spread over a wide range of subjects.
A well-known example in condensed matter physics, where the concept of
the chiral condensates or the Dirac and Haldane masses plays a central
role, is in the case of graphene in a strong magnetic
field~\cite{Khveshchenko:2001zza}, which exhibits an anomalous quantum
Hall effect~\cite{Gorbar:2008hu}.
These phenomena (and the role that the chiral condensate plays) can be
understood intuitively from the following argument: the chiral
condensate consists of 
a left-handed (or right-handed)
  fermion and a right-handed (or left-handed) antifermion
  with zero
net momenta, so the momentum directions should be opposite and the
spin directions antiparallel to each other.  Since a strong magnetic
field tends to align the spins of fermion and antifermion, it
enhances the chance for dynamical breaking of chiral symmetry to
occur.

Symmetry breaking is often modeled by an interacting low-energy
effective theory, expressed in terms of fermion degrees of freedom,
like in the BCS theory.  If we denote the coupling constant that
characterizes the strength of the interaction with $\lambda$, then
spontaneous symmetry breaking usually occurs when the coupling
constant exceeds some critical value $\lambda_c$, i.e.\
$\lambda > \lambda_c$.  When the presence of an external magnetic
field $B$ triggers the condition $\lambda_c\to 0$, we say that
``magnetic catalysis'' is realized.  This phenomenon, originally
discovered in Refs.~\cite{Klimenko:1992ch,Gusynin:1994re} (see
Ref.~\cite{magnreview} for reviews), is caused by an infrared (IR)
singular contribution to the effective potential, strengthened by the
effect of the dimensional reduction due to the magnetic field.  In
the following, the term magnetic catalysis will always be used in
this sense (and not in reference to the increasing behavior of
$\langle\bar{\psi}\psi\rangle$ as a function of increasing $B$, as it
is sometimes found in the literature~\cite{Shushpanov:1997sf}).

The magnetic catalysis is naturally lost at finite temperature, since
there are no Matsubara zero modes and therefore no IR singularity,
explaining the presence of a finite-$T$ phase transition even when
$B\neq 0$~\cite{Fukushima:2012xw}.  Interestingly, a similar mechanism
works when the spatial geometry is curved.  Such similarities and some
potential implications have recently been discussed in the context of
the ``chiral gap effect'', leading to an intuitive understanding of
the fate of chiral symmetry in the presence of nonzero
curvature~\cite{Flachi:2014jra}.  The chiral gap effect is a robust
consequence arising from the quasi-nonperturbative ($R$-resummed)
form of the propagator in curved space originally formulated in
Ref.~\cite{Parker:1984dj} (see Ref.~\cite{Flachi:2010yz} for a
discussion in relation to strongly interacting fermions and
Refs.~\cite{parkertoms, avra} for additional discussion on heat-kernel
resummations).
The situation we wish to consider in the present Letter concerns the
combined effect of curvature and magnetic fields and the question we
intend to address is whether the magnetic catalysis can be deactivated
in the presence of curvature.  As we will show, something novel
happens due to the interplay of magnetic fields and geometry.  This is
an intriguing question, leading to profound consequences for many
physical systems.  For instance, it is of relevance to astrophysical
systems (e.g., neutron stars) that involve not only gravity but also
strong magnetic fields.  In condensed matter physics, it is also
possible to induce a curvature locally by the insertion of defects in the
lattice of strongly coupled materials and use the combined effects of
geometry and external magnetic fields to probe the nature of chiral
symmetry.  Motivated by these applications, in the following we shall
focus our attention to the physically relevant case of $eB\gg R$, where $R$ indicates the 
Ricci scalar.
\vspace{0.5em}

%%%%%%%%%%   Chiral gap effect and magnetic inhibition   %%%%%%%%%%
\paragraph{Curvature induced chiral gap and magnetic inhibition.} 
It is an immediate consequence of the chiral gap effect that the IR
singularity responsible for the magnetic catalysis disappears unless
we take account of higher-order terms.  Since it will turn out to be
quite instructive, we shall explicitly check this.  The partially
resummed heat-kernel expansion reads
\begin{equation}
 \text{Tr}\, e^{-t \mathcal{D}}
 = \frac{1}{(4\pi t)^{D/2}} e^{-M_R^2 t} \frac{eB t}{\sinh(|eB|t)}\,
  e^{\frac{it}{2}eF_{\mu\nu}\sigma^{\mu\nu}}\,\sum_{k=0}^\infty a_k\,t^k \;,
\label{eq:heat}
\end{equation}
where $\mathcal{D}$ represents the Dirac operator.  

{The above expression is obtained by means of a double resummation over the scalar curvature and of the purely magnetic contributions. The first one gives rise to the first exponential factor, while the second one resums the purely magnetic contributions. The first kind of resummation has been derived in Ref.~\cite{Parker:1984dj}, while the second one has been discussed, for example, in Ref.~\cite{avramidi} (see also Chap. 5 of Ref.~\cite{parkertoms} for a nice derivation).}
The shift in the mass due to the scalar curvature resummation, i.e.\
$M_R^2\equiv M^2+R/12$, represents the essence of the curvature-induced chiral gap.  We can describe a qualitative mechanism
  for this effect by directly looking at the spectra of the Dirac
operator.  For concreteness, let us consider the case of
$D$-dimensional de Sitter space $S^D$ for which the eigenvalues of the
\textit{free} Dirac operator are \cite{Camporesi:1995fb}
\begin{equation}
 \lambda_n^{(\pm)} = m \pm i \sqrt{\frac{R}{D(D-1)}} \Bigl(
  n + \frac{D}{2} \Bigr) \;.
\label{eq:eigen}
\end{equation}
Similarly to the case of a box with periodic boundary conditions, the
spectra, labeled by an integer index $n\ge 1$, become discrete in the
present case.  For sufficiently large $D$ and taking into account the
degeneracy for each $n$, we can recover a continuum spectra (labeled by continuum momenta), 
except for the presence of a gap
proportional to $\sqrt{R}$.  This gap prevents the Dirac eigenvalues
from accumulating around the zero mode, disfavoring the formation of
a nonvanishing chiral condensate when $R>0$, as it follows
straightforwardly from the Banks-Casher relation.  In fact,
Eq.~\eqref{eq:eigen} allows for a deeper understanding of such
a curvature-induced chiral gap: If we take the product
$\lambda_n^{(+)}\lambda_n^{(-)}$, then the leading effect of $R$
naively looks like a shift in $m^2$ by $R$, but in the original
eigenvalues $\lambda_n^{(\pm)}$ of the first-order Dirac operator it
is obvious that this $R$-induced shift is not on the real axis, but it
occurs along the imaginary axis, which explains how such a curvature-generated mass gap can be consistent with chiral symmetry.

Picking up the first contribution with $a_0=1$ from
Eq.~\eqref{eq:heat}, we can express the one-loop effective potential as
\begin{equation}
 \begin{split}
 &V_{\text{$R$-resum}}[M] = \frac{M^2}{2\lambda_D}\\
 &\qquad + \kappa_D (eB)
  \int_{1/\Lambda^2}^\infty dt\, t^{-D/2}\, e^{-M_R^2 t}\, \coth(eB t) \;,
 \end{split}
\label{eq:poten_R}
\end{equation}
where we have used proper-time regularization.  In the above
expression, we have denoted the coupling constant in $D$ dimensions
with $\lambda_D$ and defined
$\kappa_D\equiv 2^{\lfloor(D+1)/2\rfloor-1}/(4\pi)^{D/2}$.  In
obtaining Eq.~\eqref{eq:poten_R}, we could have adopted any other
regularization scheme ($\zeta$ function, Pauli-Villars, etc.) as long
as gauge invariance is preserved.  Different prescriptions would not
change the IR singular structure.  In the limit of
$eB\gg\Lambda^2$, $\coth(eB t)\approx 1$, and, for $R=0$, we can
approximate the potential as
\begin{equation}
 V_{\text{catalysis}}[M] \approx \frac{M^2}{2\lambda_4}
  + \kappa_4 (eB)M^2 \biggl( -1 + \gamma_{\text{E}}
  + \ln\frac{M^2}{\Lambda^2} \biggr)
\end{equation}
for $D=4$.  We may then define a ``renormalized'' coupling as
$1/(2\lambda_4')\equiv 1/(2\lambda_4) + \gamma_{\text{E}}\kappa_4 (eB)$,
and observe that the dynamical mass always takes a finite value:
$M^2=\Lambda^2 \exp[-1/2\kappa_4 \lambda_4' (eB)]$.  This is how the
magnetic catalysis works.

It should be noted that the IR singularity possibly remains if $R<0$
(see Ref.~\cite{Gorbar:1999wa} for the analysis with $R<0$).  Then it
is natural to expect that effects of negative curvature may be
  compensated by those of temperature leading to a restoration of the catalysis.
For $R>0$, however, $M_R^2$ never reaches zero and
$\ln(M_R^2/\Lambda^2)$ is no longer IR singular, which could be
regarded as a mechanism of ``magnetic inhibition'' induced
geometrically (having an origin totally different from the
  inhibition at finite temperature~\cite{Fukushima:2012kc}).
\vspace{0.5em}

%%%%%%%%%%   Magnetic catalysis from higher-order terms   %%%%%%%%%%
\paragraph{Magnetic catalysis from higher-order terms.} 
The above is not the whole story.  As done in the derivation of the
chiral gap effect, below we shall assume a geometrical structure such
that we can neglect Ricci and Riemann tensors as compared to
the Ricci scalar: $|R_{\mu\nu}|\ll |R|$ and
$|R_{\mu\nu\rho\sigma}| \ll |R|$, as it happens, for instance, for
maximally symmetric geometries with large $D$.  However, in the
present case, since higher-order contributions coming from $a_k$ with
$k\ge2$ involve terms that mix the magnetic field with curvature,
these cannot be discarded, but may dominate over the purely
gravitational tensorial combinations.
Before proceeding with detailed calculations including higher-order
terms, it is useful to obtain a parametric form for the chiral
condensate or the dynamical mass (correction) by using dimensional
analysis. Let us suppose that higher-order terms give rise to
an IR-singular correction $\delta M^2$ to the dynamical mass.  Then
$(eB)^2R$ would be the most natural combination from which the correct
mass dimensionality arises.  This is because $a_k$ should vanish
(except for subleading contributions involving
$R_{\mu\nu}$ and $R_{\mu\nu\rho\sigma}$) when we take either $B=0$ or
$R=0$.  Since $B$ is generated from a Lorentz-invariant contraction of
$F_{\mu\nu}$, the lowest order should be $(eB)^2$.  Finally,
multiplying by the four-fermion coupling constant $\lambda_D$ that has
a mass dimension, we conclude that $\delta M^2$ should be a function
of $\lambda_D(eB)^2R$.

In $D$-dimensional space-time, the mass dimension of $\lambda_D$ is
$[\lambda_D]=2-D$, and so $[\lambda_D (eB)^2 R]=8-D$.  Thus,
dimensional analysis allows us to write
\begin{equation}
 \delta M^2 \;\sim\; \bigl[ \lambda_D (eB)^2 R
  \bigr]^{2/(8-D)} \;.
\label{eq:dM}
\end{equation}
The above expression indicates that $D=8$ is a critical dimension.
For the $D=8$ case, $\lambda_8 (eB)^2 R$ becomes dimensionless, and
the parametric dependence of $\delta M^2$ should become logarithmic,
i.e., $\delta M^2\sim \Lambda^2 \exp[-C/\lambda_8 (eB)^2 R]$, which
reminds us of the standard magnetic catalysis.

Let us now look into the concrete calculations of the chiral
condensate and of the effective potential.  Since we are working in
the regime where $eB\gg R$, in the heat-kernel expansion we need to
resum all the terms of the form $(eB)^2 R\, (eB t)^k$ with $k\ge 2$.
Here, we adopt the same strategy as when performing the
$R$ resummation that leads to an exponential dependence.  Namely, we
postulate a resummed expansion based on the following reorganisation
of the various terms:
\begin{equation}
 \sum_{k=0}^\infty a_k\,t^k
 = 1 + R \sum_{k=1}^\infty \alpha_k\, (eB)^{k-1} t^k\, e^{\beta_k eBt} \;,
\end{equation}
where $\alpha_k$ and $\beta_k$ are dimensionless coefficients {that can be computed explicitly from the heat-kernel coefficients. 
The above expression can be obtained by direct construction of the corrections to the zeta function the higher-order mixed terms induce (the first few coefficients can be explicitly obtained from those reported in Ref.~\cite{Parker:1984dj} by keeping only the terms that mix the magnetic field with curvature), by ordering according to their mass dimensionality and exponentiating.} 

We explicitly obtained $a_2$ and $a_3$ ($a_1=0$ from an obvious reason
of dimensionality) and identified the coefficients $\alpha_2$ and
$\beta_2$ (and $\alpha_1=0$ corresponding to $a_1=0$).  Similar
computations of heat-kernel coefficients are reported in
Refs.~\cite{Parker:1984dj,avramidi}.  After lengthy calculations, we
obtained the effective potential that reads
\begin{align}
 &V_{\text{$B$-resum}}[M] = \frac{M^2}{2\lambda_D} + \kappa_D (eB)
  \int_{1/\Lambda^2}^\infty dt\, t^{-D/2}\, e^{-M_R^2 t} \notag\\
 &\qquad\qquad
  + \frac{\kappa_D (eB)^2 R}{6D(D-1)} \int_{1/\Lambda^2}^\infty dt\,
  t^{-D/2+2}\, e^{-M_B^2 t} \;.
\label{eq:poten}
\end{align}
Since we are interested in whether the magnetic catalysis occurs in
the presence of curvature
when $eB$ is the largest scale in the system, we have approximated
$\coth(eBt)\approx 1$ in the above expression.  We defined
$M_B^2\equiv M_R^2 + [2(4-D)/15] eB$, which represents a $B$-induced
correction to the chiral gap effect.  This expression implies that the
curvature-induced chiral gap $R/12$ could be compensated by the effect
of the magnetic field for $D-4> 0$ if $eB\gg R$.

{The last term in Eq.~(\ref{eq:poten}) represents the correction to the potential coming from higher-order terms mixing curvature tensors with magnetic field and that can be obtained, at the price of a long calculation, by starting from the general expression of the zeta function and from the explicit knowledge of the heat-kernel coefficients.}

The term that carries the potential IR singularity is the third term
in the effective potential~\eqref{eq:poten}.  This can be explicitly
evaluated, leading to
\begin{equation}
 \int_{1/\Lambda^2}^\infty dt\, t^{-D/2+2}\,e^{-M_B^2 t}
  = M_B^{D-6} \Gamma(3-D/2,M_B^2/\Lambda^2) \;,
\label{eq:third}
\end{equation}
from which it is easy to estimate the parametric dependence of the
chiral condensate, which turns out to be consistent with
Eq.~\eqref{eq:dM}.  In what follows below let us look at
Eq.~\eqref{eq:dM} for specific choices of $D$.
\vspace{0.5em}

%-----   D=3   -----%
\noindent
($D=3$).~~
This is probably the most relevant to systems in condensed matter
physics.  We should remark here that, while the condition of large $D$
allows for a simple hierarchy between curvature invariants, it is not
a necessary condition for the validity of the above statements.  In
this case, interestingly, the magnetic shift and the curvature-induced
chiral gap accumulate.  Because the correction~\eqref{eq:third} is
infrared singular as $\propto 1/M_B^3$, it would be dominant near
$M_B\sim 0$, and we can easily solve the gap equation to find
\begin{equation}
 M_B^2 = M^2 + \frac{R}{12} + \frac{2}{15}eB
  = \biggl[ \frac{\sqrt{\pi}\kappa_3 \lambda_3 (eB)^2 R}{24}
  \biggr]^{2/5} \;,
\end{equation}
which is of course consistent with Eq.~\eqref{eq:dM}.  Under the
condition of $eB\gg R$, thus, the magnetic shift term $\sim eB$ would
oversaturate $M_B^2$.  This means that there is no stable solution
near $M_B^2\sim 0$ with $M^2>0$, and $M^2$ is actually determined by
the balance with the second term in the potential~\eqref{eq:poten}
rather than the third term.  In summary,  for $D=3$, the chiral gap
effect overwhelms the magnetic catalysis, and higher-order terms would
not override this situation.
\vspace{0.5em}

%-----   D=4   -----%
\noindent
($D=4$).~~
Since there is no $B$-induced correction to the mass squared, $D=4$ is
an exceptional (and also realistic in relativistic systems) case.  It
is obvious from Eq.~\eqref{eq:third} that a singular term
$\propto 1/M_B^2$ appears in the effective potential.  Then we can
easily locate the minimum of the effective potential at
\begin{equation}
 M_B^2 = M^2 + \frac{R}{12}
 = \sqrt{\frac{\lambda_4 \kappa_4 (eB)^2 R}{36}} \;.
\end{equation}
We note that $M^2$ is nonvanishing if
$(eB)^2 > R/(4\lambda_4\kappa_4)$, which means that the magnetic
catalysis is recovered in this case for $D=4$.  Strictly speaking, for
arbitrary value of $B$, magnetic catalysis is not necessarily
guaranteed, since this inequality also determines a critical
$\lambda_4$.  We should, however, note that we postulated $eB\gg R$ in
our analysis; therefore, it is obvious that the term $R/12$ turns out
to be subdominant.

It is quite natural that $M^2$ 
is proportional to the density of states of the Landau levels,
$\sim eB$, although a rather unconventional combination of
$\lambda_4 R$ appears in addition.  Because $\lambda_4 R$ is
dimensionless in $D=4$, the dependence on $\lambda_4 R$ could have any
functional form, in principle, as long as it vanishes for $R\to 0$.
The square root dependence as seen in Eq.~\eqref{eq:dM} is very
interesting, because the right-hand side $\sim eB\sqrt{\lambda_4 R}$
could dominate over the chiral gap effect contribution $\sim R$ not
only for $eB\gg R$ but also for $R\ll \lambda_4^{-1}$ (and $B\neq 0$).
The latter is usually the case (and $\lambda_4^{-1/2}$ should be
interpreted as a typical scale in theory,
e.g.,\ $\sim\Lambda_{\rm QCD}$ in the case of quarks).
\vspace{0.5em}

%-----   D=5   -----%
\noindent
($D=5$).~~
The IR singularity is weakened as $\sim 1/M_B$ in
Eq.~\eqref{eq:third}, but it is still sufficient to lead to a chiral
condensate.  In this case we find that $M^2$ is affected by a
$B$-induced term:
\begin{equation}
 M_B^2 = M^2 + \frac{R}{12} - \frac{2eB}{15}
  = \biggl(\frac{\sqrt{\pi}\lambda_5\kappa_5 (eB)^2R}{120}\biggr)^{2/3} \;.
\label{eq:D5}
\end{equation}
Interpreting this expression requires care, particularly due to the
presence of the third term $\sim eB$ in the left-hand side.
Equation~\eqref{eq:D5} may imply that $M^2 \sim eB$ (regardless of
$\lambda_5$!) even in the limit of vanishing $R$.  Going back to
Eq.~\eqref{eq:poten}, however, the third term in the effective
potential is vanishing for $R=0$, and so $M^2$ cannot have any such
correction.  Thus, the behavior of the effective potential itself
smoothly changes from $R=0$ to $R\neq0$, but the position of the minimum
discontinuously jumps once $B$ is switched on.  This is not so unusual
if the potential shape is shallow enough. In principle, such a 
discontinuous behavior at $R=0$ would be made milder by further resummation of IR singular terms around $M_B^2\sim 0$.

We should emphasise that this ``geometrically induced magnetic
catalysis'' is absent in flat space and can occur even in the
case of a weakly curved geometry.  For this reason, its effect should
be more easily observable than the chiral gap effect itself.  In fact,
even though the correction to the potential height is negligibly small
due to suppression by $R$, a sizable shift in $M^2$ is possible.  We
did not include the IR-safe terms in the above discussion, since they,
obviously, do not change our qualitative conclusion.
\vspace{0.5em}

%-----   D=8   -----%
\noindent
($D=8$).~~
Qualitatively, the cases $D=5$, $6$, and $7$ are similar.  So, let us
consider the next nontrivial case, $D=8$.  As we have already
mentioned, $\lambda_8 (eB)^2 R$ is dimensionless in this case, and
explicit calculations lead to
\begin{equation}
 M_B^2 = M^2 + \frac{R}{12} - \frac{16eB}{15}
  = \Lambda^2 \exp\biggl[-\frac{168}{\lambda_8' \kappa_8 (eB)^2 R}
  \biggr] \;,
\end{equation}
where, for notational convenience, we have defined
$\lambda_8'^{-1}\equiv \lambda_8^{-1}
+(-1+\gamma_{\rm E})\kappa_8 (eB)^2 R/168$.

Although this nonanalytic form in terms of the coupling $\lambda_8'$
is peculiar, physical consequences are dominated by a shift of
$-16eB/15$ in $M_B^2$ that induces $M^2\sim eB$ as long as $M_B^2$ is
non-negative.  The same argument can be applied to larger $D$.  Hence,
regarding the behavior of $M^2$, we can generalize our study for any
$D>4$ and conclude that the essential feature of the geometrically
induced magnetic catalysis should be common if $D$ is larger than
  the critical dimension, $D=4$.
\vspace{0.5em}

%%%%%%%%%%   Conclusions and outlooks   %%%%%%%%%%
\paragraph{Conclusions and outlook.} 
In this Letter, we have investigated the two competing effects of
curvature and magnetic field with the intent to clarify the interplay
between the chiral gap effect and the magnetic catalysis.  In
conformity with the chiral gap effect, quasi-nonperturbative
contributions (due to $R$ resummation in the heat-kernel) induce a
curvature correction to the mass and regulate, as it may be
intuitively expected, the infrared singularity deactivating the strict
magnetic catalysis at this level of approximation.

However, we have discovered (using both dimensional analysis and by
means of explicit evaluation of the effective potential) that $D=4$ is
a critical dimension.  Above $D=4$, next-to-leading-order corrections
in the heat-kernel expansion become relevant, restoring the magnetic
catalysis in a novel, geometrically induced fashion.  We have also
seen that below $D=4$ these corrections become irrelevant for the
magnetic catalysis, and the curvature and magnetic field cooperate.  More
precisely, for $D>4$, the infrared singularity induces a dynamical
mass of the order of $eB$, though the singular contribution to the
effective potential is suppressed by small $R$.  In the critical $D=4$
case, the dynamical mass squared is proportional to
$eB \sqrt{\lambda_4 R}$, where $\lambda_4$ is the four-fermion coupling
constant and the combination $\lambda_4 R$ is dimensionless.

Possible applications of the geometrically induced magnetic catalysis
discussed here are of relevance for a wide variety of physical
setups.  Perhaps the most natural environment is that of neutron
stars where both a curvature and a magnetic field (that could
  be as strong as $10^{12}\,{\rm T}$~\cite{Makishima:2014dua})
could lead to a realization of what we have discussed here.
Micro black holes in the early Universe or in  
high-energy particle collisions accelerator experiment could also be an interesting playground for
  the chiral gap effect and the geometrically induced magnetic
  catalysis.
In ultrarelativistic nucleus-nucleus collision, a
  pulsed and strong magnetic field is generated, and the space-time
  geometry is nontrivial due to expansion and flowing fluids  (that
  may cause horizon formation or an acoustic metric; see
  Refs.~\cite{Kharzeev:2005iz,Labun:2010wf} for related discussions).
Finally, although for lower dimensionality the magnetic field and the
curvature cooperate (rather than compete), the present discussion may
be of relevance in the context of strongly coupled layered materials,
where curvature can be generated locally by the insertion of defects.

In order to be able to discuss the problem in a 
model-independent way, in this work we have focused only on the
strict characterization of the magnetic catalysis.  It would be
intriguing to use chiral perturbation theory, that is, a theoretical
approach complementary to the fermionic description adopted here.
Also, studying the behavior of the chiral condensate as a
  function of various external parameters including the temperature
  effect would be interesting.  These remain as future problems.
\vspace{0.5em}

\acknowledgments
This work was partially supported by JSPS KAKENHI Grant No.
24740169 (K.F.), by the Funda\c{c}\~{a}o para a
Ci\^{e}ncia e a Tecnologia of Portugal (FCT) and the Marie Curie
Action COFUND of the European Union Seventh Framework Program Grant
Agreement No. PCOFUND-GA-2009-246542 (A.F.), and by the Funda\c{c}\~{a}o para a
Ci\^{e}ncia e a Tecnologia of Portugal (FCT) through Grant No. SFRH/BPD/77678/2011 (V.V.).

\end{document}